\documentstyle[11pt,twocolumn,psfig]{nature}
\def\be{\begin{equation}}
\def\ee{\end{equation}}

\def\Msun{{M_\odot}}
\def\Zsun{{Z_\odot}}

\def\gsim{\lower.5ex\hbox{\gtsima}}
\def\lsim{\lower.5ex\hbox{\ltsima}}
\def\gtsima{$\; \buildrel > \over \sim \;$}
\def\ltsima{$\; \buildrel < \over \sim \;$}
\def\prosima{$\; \buildrel \propto \over \sim \;$}
\def\gsim{\lower.5ex\hbox{\gtsima}}
\def\lsim{\lower.5ex\hbox{\ltsima}}
\def\simgt{\lower.5ex\hbox{\gtsima}}
\def\simlt{\lower.5ex\hbox{\ltsima}}
\def\simpr{\lower.5ex\hbox{\prosima}}

\def\beq#1{\begin{equation}\label{#1}}
\def\eeq{\end{equation}}
\def\beqa#1{\begin{eqnarray}\label{#1}}
\def\eeqa{\end{eqnarray}}

\def\H2p{H$_2^+$ }

\def\mH2p{H_2^+}

\title{
A SUPERNOVA ORIGIN
FOR DUST IN A HIGH-REDSHIFT QUASAR
}

\mainauthor{R. Maiolino et al.}
\author{{
R. Maiolino\affiliation{INAF - Osservatorio Astrofisico di
Arcetri, Largo Enrico Fermi 5, 50125 Firenze, Italy}, 
R. Schneider\affiliation{``Enrico Fermi'' Center, Via Panisperna 89/A, 00184 Roma, Italy}$^*$, 
E. Oliva\affiliation{Telescopio Nazionale Galileo,
 C. Alvarez de Abreu, 70, 38700 S.ta Cruz de La Palma, Spain}$^*$, 
S. Bianchi\affiliation{CNR-IRA, Sezione di Firenze, Largo Enrico Fermi 5,
50125 Firenze, Italy},\\ 
A. Ferrara\affiliation{SISSA/International School for Advanced Studies,
	   Via Beirut 4, 34100 Trieste, Italy},
F. Mannucci$^{\S}$,
M. Pedani$^{\ddag}$,
M. Roca Sogorb\affiliation{Astrofisico Fco. S{\`a}nchez, Universidad de La Laguna, 38206 La Laguna, Tenerife, Spain} 
}}

\begin{document}
\summary{
Interstellar dust plays a crucial role in the evolution of the Universe by
assisting the formation of molecules\cite{hirashita02}, by triggering the formation of the
first low-mass stars\cite{schneider03}, and by absorbing stellar ultraviolet-optical light
and subsequently re-emitting it at infrared/millimetre wavelengths.
Dust is thought to be produced predominantly in the envelopes of evolved (age
$>$1 Gyr), low-mass stars\cite{whittet}. This picture has, however, recently
been brought into question by the discovery of large masses of dust in the host
galaxies of quasars\cite{bertoldi03,priddey03} at redshift $z>6$,
when the age of the Universe was less than 1 Gyr.  Theoretical
studies\cite{todini01,nozawa03,schneider04}, corroborated by
observations of nearby supernova
remnants\cite{moseley89,dunne03,morgan03}, have suggested that supernovae
provide a fast and efficient dust formation environment in the early
Universe. Here we report infrared observations of a quasar at
redshift 6.2, which are used to obtain directly its dust extinction curve.
We then show that such a curve is in excellent agreement with supernova dust
models. This result demonstrates a supernova origin for dust in this
high-redshift quasar, from which we infer
that most of the dust at high redshifts has probably the same origin.
}

\maketitle

\bigskip
\noindent

Powerful quasars offer an optimal tool for detailed studies of dust 
in their host galaxies out to very high redshifts. 
Dust extinction is inferred through reddening 
of the quasar UV/optical continuum emission.
The extinction curve ($A_{\lambda}=1.086~\tau_{\lambda}$,
where $\tau_{\lambda}$ is the wavelength-dependent optical depth) of dust associated with 
low redshift, mildly obscured
quasars has been typically found to be consistent with that of  
the Small Magellanic Cloud (SMC)\cite{richards03,reichard03,hopkins04},
while for heavily absorbed quasars there
are indications that the extinction curve may be
different\cite{gaskell04,maiolino01}. 

An important class of quasars are the
Broad Absorption Line (BAL) quasars, whose UV spectrum
is characterized by blueshifted,
deep and broad absorption features associated with highly ionized 
atomic species,
which trace powerful outflows of dense gas along our line
of sight.
Low Ionization BAL (LoBAL) quasars are characterized by additional
low ionization absorption lines, probably associated 
with higher column densities of gas.
LoBAL quasars at $z<4$ are {\it always} significantly reddened
by dust (associated with the outflowing gas),
and therefore they are ideal laboratories to
investigate the properties of dust.

By means of low resolution
near-IR spectroscopic observations
we have recently identified a few BAL quasars at $z\approx5-6$
(Maiolino et al. 2004).
The most distant among them is the LoBAL quasar SDSSJ104845.05+463718.3 
(hereafter SDSS1048+46). At a redshift $z=6.193$ (based on
a new medium resolution spectrum around MgII$\lambda$2798; Maiolino 
et al. in prep), this is the only LoBAL at $z>5$ currently known, and
offers a unique chance of investigating the dust extinction curve 
at $z\approx6$.
This quasar was re-observed at higher spectral resolution with the goal of
de-blending
the deep troughs characterizing its spectrum and of estimating the reddening
of the continuum. Observations were obtained in February 2004 with the
Near Infrared Camera Spectrometer (NICS\cite{baffa01}) at the
Telescopio Nazionale Galileo (Canary Islands).
SDSS1048+46 was observed for a total of 3 hours,
with a slit of 1.5$''$ oriented along the parallactic angle.
We used a grism covering the spectral region $0.8-1.45\mu$m.

Fig.~1a shows the observed spectrum
(solid line) smoothed to a lower resolution for
sake of clarity, and combined with a previous low resolution spectrum\cite{maiolino04}
of the full wavelength range $0.8-2.4\mu$m. At the quasar's redshift our data cover the
wavelength range $1200<\lambda _{rest} <3300$\AA \ in the rest frame of
the emitted radiation.
The observed spectrum is much bluer than LoBAL quasars observed
at $z<$4, whose average spectrum is shown with the dashed
line\cite{reichard03}.
In particular, SDSS1048+46 is bluer than {\it any} known
LoBAL quasar\cite{reichard03} at $z<$4.
Since the red shape of LoBAL quasars
at z$<$4 is ascribed to dust reddening,
the much bluer slope of SDSS1048 already
indicates a large variation of dust extinction 
and reddening from $z<4$ to $z\approx6$.

The continuum at
$\lambda_{rest}>$1700~\AA \ can be fitted with a non-BAL (unreddened)
quasar template with a slope  $\alpha=-2.1$ ($F_{\lambda}\propto\lambda^{\alpha}$), which is 
shown with a dotted line in Fig.1. Such a blue slope is consistent with the
intrinsic slope of --2.01 inferred by Reichard et al. (2003)\cite{reichard03}
for BAL quasars.
The observed spectrum deviates significantly
from the non-BAL unreddened template only at $\lambda_{rest}<1700~$\AA.
While such a deviation was initially ascribed to a severe blend of the CIV and
SiIV troughs in the low resolution spectrum, the new medium resolution
spectrum (whose unsmoothed and enlarged version is shown in Fig.~1b) clearly
shows that this  is not the case. The continuum outside the troughs
is indeed redder than expected by the extrapolation of the longer
wavelength spectrum through the unreddened quasar template
(dotted line). If such reddening at short wavelengths is due to dust,
then the extinction curve must be quite unusual: relatively flat at
$\lambda>1700$~\AA \ and steeply rising at shorter wavelengths.
We can quantitatively derive the extinction curve by using the equation
$A_{\lambda}=-2.5~log(F_{obs}/F_{intr})$, where $F_{obs}$
is the observed spectrum and $F_{intr}$ is the intrinsic spectrum.
We avoided spectral regions contaminated by emission/absorption features
by interpolating the continuum from the nearby regions.
The blend of FeII lines\cite{maiolino03} makes more
uncertain the description
of the continuum in the spectral region between 2300\AA \ and 3050\AA;
however, the latter spectral region is not critical within
the context of this paper, since the most important aspect is that the
extinction curve must be rather flat in the region between 1700\AA \  and
3300\AA, and increase rapidly at $\lambda _{rest} < 1700$\AA.
For the intrinsic spectrum we used the non-BAL template
obtained by the SLOAN survey\cite{reichard03} 
(this choice is appropriate since non-BAL quasars at z$\sim$6 have
spectra similar to those at z$<$4)\cite{maiolino04}, with
slopes ranging from $\alpha=-2.1$ (the reddest slope compatible
with our observed spectrum at $\lambda _{rest}>1700$\AA)
to $\alpha=-2.5$ (only 1\% of quasars 
have slopes bluer than this value\cite{reichard03}).
The resulting
extinction curve is shown in Fig.~2 (thick solid line and
shaded region).
As expected, the extinction curve inferred for this quasar at $z=$6.2
is quite different with respect to the SMC curve which applies
to quasars at $z<4$.
The dot-dashed line in Fig.~1b indicates
the non-BAL template absorbed with the extinction curve inferred
by us in Fig.2, which nicely matches the observed spectrum
(except for the emission and absorption lines, whose intensity and shape
depend on the physics of the ionized gas). The extinction
$A_{\mbox{3000\AA}}$ required
to match the observed spectrum is in the range 0.4-0.8 mag.

The extinction curve inferred from the most distant LoBAL can also reproduce the
shape of the second most distant BAL SDSS1044-01, a High Ionization BAL
(HiBAL)\cite{maiolino04} at $z=$5.78.
The latter quasar presents much lower reddening
than SDSS1048-46 and therefore cannot be used to provide tight constraints on the
extinction curve. Nonetheless, similarly to SDSS1048-46, it is characterized a rather blue
continuum at $\lambda _{rest}>1700$\AA \ and a reddening
at $\lambda _{rest}<1700$\AA ,
which can be nicely fitted by using the same extinction curve derived above and shown
in Fig.~2.

In order to interpret the observations, we have computed the SN dust 
extinction curve by
using the model proposed by Todini \& Ferrara (2001)\cite{todini01} 
which describes dust formation in the ejecta of Type-II SNe as a function of the
progenitor mass and metallicity. 
In spite of the uncertainties, the model has been successfully applied to 
interpret the observed properties of SN 1987A\cite{todini01} and of the young, metal-poor 
dwarf galaxy SBS 0335-052\cite{hirashita02b}. 
A grid of SN models has been
considered, with different initial stellar progenitor masses
($12~\Msun\leq M\leq40~\Msun$) 
and metallicities ($0\leq Z/\Zsun\leq1$)\cite{woosley95}.
The resulting SN dust grains are made of silicates (Mg$_2$SiO$_4$ and MgSiO$_3$), amorphous carbon (AC),
magnetite (Fe$_3$O$_4$) and corundum (Al$_2$O$_3$).
The typical grain sizes range between 5~\AA~to 0.1 $\mu$m,
with smaller grains being predominantly composed
of silicates and magnetite (sizes $\approx $5--15~\AA)
and larger ones by amorphous carbon 
(sizes $\approx 100-200$~\AA).
The predicted grain properties for different
Type-II SNe  have
then been averaged over the stellar Initial Mass Function (IMF).
We adopted a Salpeter IMF, but even higher-mass weighted
Larson-like IMFs\cite{larson98} do not change significantly the
results ($\simlt$2\% in the extinction curve).

The SN dust extinction properties have been derived using the standard Mie
theory for spherical grains. Grains made of AC are the main contributors
to extinction. We used the optical properties for AC produced in an inert
atmosphere ({\em ACAR}\cite{zubko96}), which is appropriate for the
SN ejecta environment.
Other important contributors to extinction were
found to be Mg$_2$SiO$_4$ and Fe$_3$O$_4$ grains\cite{scott96,mukai89}.

Fig.2 shows the resulting extinction curves, which
are in a much better
agreement with the observations than the usual SMC curve.
The extinction curves were obtained for various initial stellar metallicities.
The best agreement is found for the $Z=10^{-2}~Z_{\odot}$ and
$10^{-4}~Z_{\odot}$ models,
although the other models do not differ strongly and are still
within the observational uncertainties.
We also show the case of  dust formed
in the ejecta of a single 25 $\Msun$, $Z=10^{-4}Z_{\odot}$ Type-II SN;
the agreement for this case is excellent at most wavelenghts.
The plateau between 1700 \AA~and 3000 \AA~is due to a local
minimum between two broad absorption features caused by
AC grains, with the added flat contribution from Fe$_3$O$_4$
grains. Silicate grains contribute to the rise at shorter wavelenghts.

We conclude that our analysis, combined with observations,
provides the first direct evidence for dust produced in
SN ejecta, rather than in evolved stars, in an object at $z>6$.
In particular, we
find that dust purely produced by Type-II SNe can explain
very well the observed extinction curve.
By implication, we conclude that much of the dust seen at high redshifts
probably has the same origin.
The properties
of high redshift dust are therefore noticeably different from those found at later cosmic
times:
grains are typically smaller, due to their different formation history and to the short time
available to subsequently accrete heavy atoms and coagulate
with other grains.

\bigskip
\noindent
{\bf Correspondence and requests for materials should be addressed
to Roberto Maiolino
(e-mail: maiolino@arcetri.astro.it).}

\bigskip
\noindent
{\bf Acknowledgements}
All authors have contributed equally to this paper.
We are grateful to J. Brucato for providing the
optical constant of ACAR grains and we are grateful
to M. Walmsley for useful comments.
This work was partially supported by the Italian
Ministry of Research (MIUR) 
and by the National Institute for Astrophysics (INAF).

\bigskip
\noindent
{\bf Competing interests statement} The authors declare that they
have no competing financial interests.

\newpage

%\vskip0.5cm
%\noindent
%{\bf Figure legends:}

\vfill
\eject

\begin{figure*}
%\centerline{\psfig{figure=figure1.eps,width=10cm}}
\vskip-0.5cm
\hskip3cm
\psfig{figure=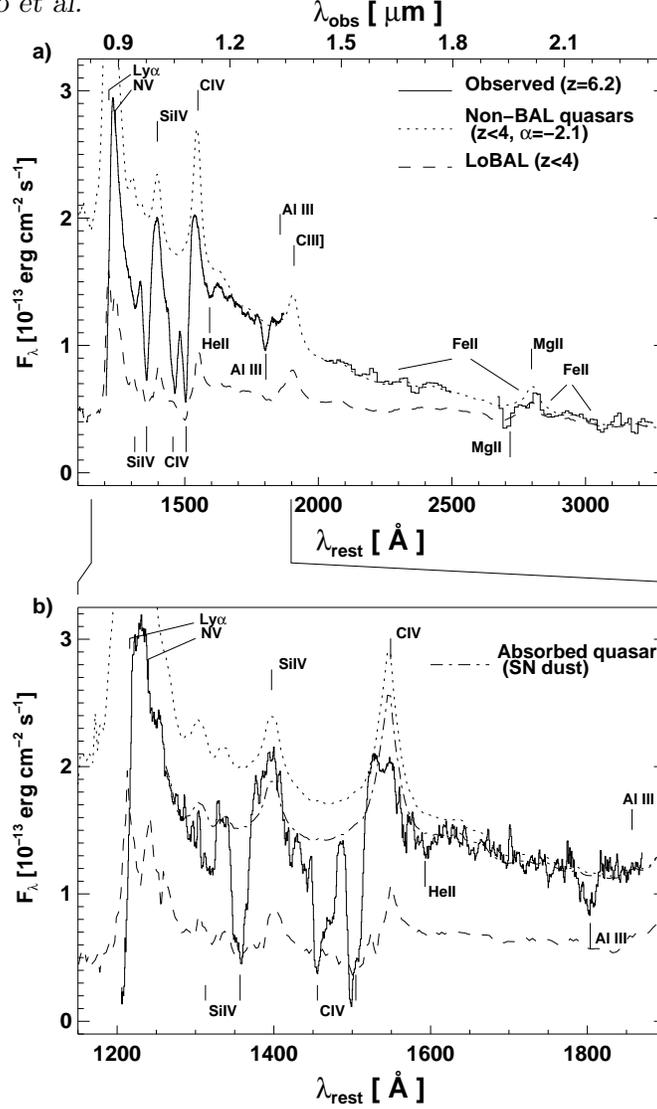,width=8.7cm}
\vskip-0.5cm
\caption{\footnotesize
%\noindent
%{\bf Figure 1.}
Spectrum of the quasar SDSS1048+46 at redshift 6.2, compared with quasar
templates at lower redshift.
{\sl a)} The solid
line shows the low resolution near-IR spectrum 
of the Low Ionization Broad Absorption Line (LoBAL) quasar SDSS1048+46.
It is the composition of a previous low resolution $\lambda /\Delta \lambda = 75$
spectrum\cite{maiolino04} (shown only for $\lambda > 1.4\mu m$),
and the new medium
resolution $\lambda /\Delta \lambda = 350$ spectrum smoothed to a resolution
of 75 for sake of clarity.
Cross-calibration between the two
spectra is ensured by matching the common parts.
Upper labels identify the emission lines
while lower labels identify the blueshifted absorption features.
The presence of strong absorption from the high ionization
species CIV$\lambda$1549 and SIV$\lambda$1397,
along with absorption by the low ionization species AlIII$\lambda$1857 and
MgII$\lambda$2798 classifies this as a LoBAL quasar.
The missing parts of the spectrum are due to the regions of bad
atmospheric transmission.
The dotted line is the average spectrum of (unreddened)
non-BAL quasars at $z<4$ with a slope $\alpha=-2.1$.
The dashed line is the average spectrum of LoBAL quasars at
$z<4$.
{\sl b)} Same as above, but where the solid line
shows only the (unsmoothed) medium resolution spectrum
($\lambda /\Delta \lambda = 350$)
of SDSS1048+46. It is important to note the continuum level
outside the troughs at $\lambda<1700$\AA,
which is well below the extrapolation of the continuum
at longer wavelenghts in the case of no dust reddening (dotted line).
Both low and high resolution spectra allow the identification of the
spectral regions for the continuum fitting (required to derive
the extinction curve), and more specifically:
1275--1290\AA, 1325--1340\AA, 1417--1434\AA, 1482--1489\AA,
1577--1590\AA, 1567--1677\AA, 1708--1782\AA, 1815--1845\AA,
2015--2270\AA, 3050--3255\AA.
The dot--dashed line in panel {\sl b)}
shows the effect of absorbing the non-BAL template
with the extinction curve inferred by us and shown in Fig.~2 (thick solid
line).}
\end{figure*}

\vfill
\eject

\begin{figure*}
%\centerline{\psfig{figure=figure2.eps,width=10cm}}
\hskip2cm
\psfig{figure=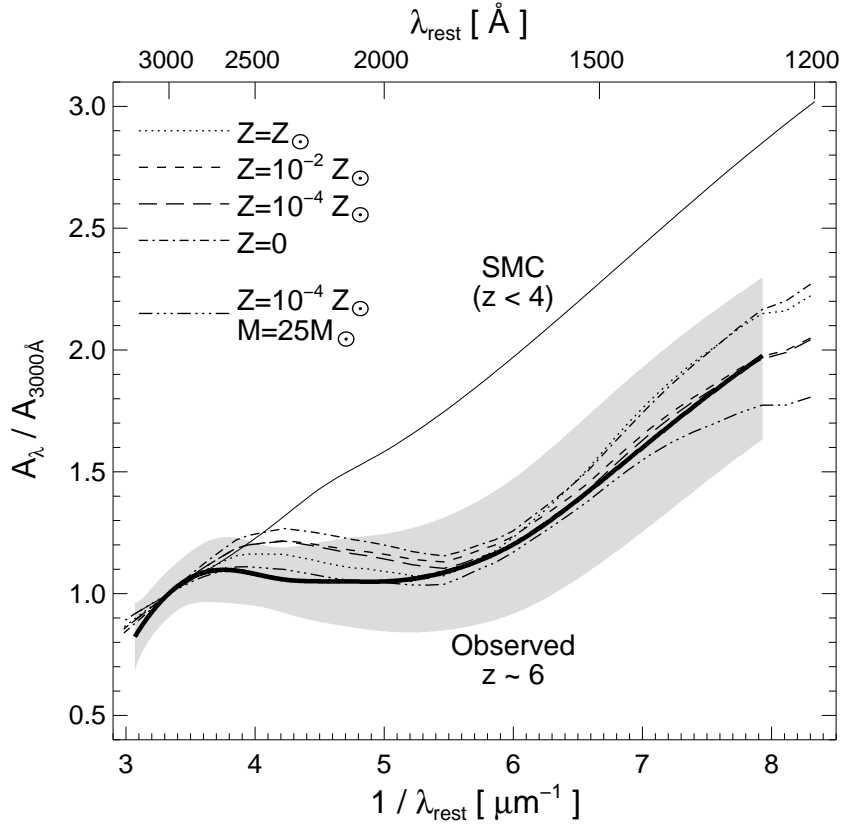,width=11cm}
\caption{\footnotesize
%{\bf Figure 2.}
Extinction curve observed in the quasar SDSS1048+46 at $z=6.2$ compared
with the extinction curve observed in quasars at z$<$4
and with the extinction curve expected from supernova (SN) dust.
The thick solid line shows
the extinction curve inferred for SDSS1048+46. The shaded
area shows the associated uncertainty,
which includes the range of intrinsic spectral slopes
(this uncertainty dominates at short wavelenghts, $\lambda < 1700$\AA),
the uncertanty on the continuum interpolation (in particular around
the FeII hump between 2300\AA \ and 3050\AA), and the noise in the
spectrum.
The thin solid line
shows the Small Magellanic Cloud (SMC) extinction curve,
which applies to quasars at $z<4$ (including BAL quasars).
The other lines indicate theoretical
predictions for dust produced by SNe. More specifically:
dot-dashed, long-dashed, short-dashed and dotted lines represent 
the extinction curves produced by SNe obtained assuming that stars
form according to a Salpeter IMF and initial 
metallicities $0$, $10^{-4}$, $10^{-2}$, and 1~$Z_{\odot}$, respectively.
We also show the case of  dust formed 
in the ejecta of a single 25~$\Msun$, $Z=10^{-4}~Z_{\odot}$ Type-II SN
(triple-dot-dashed line).}
\end{figure*}

\end{document}